\RequirePackage[l2tabu, orthodox]{nag}
\documentclass[11pt]{article}

\usepackage[T1]{fontenc}
\usepackage{tcolorbox}

\usepackage{soul} 

\usepackage{garamond}  
\renewcommand{\rmdefault}{ugm}

\usepackage[bitstream-charter]{mathdesign}
\usepackage{amsmath}
\usepackage[scaled=0.92]{PTSans}

\usepackage[
  paper  = letterpaper,
  left   = 1.65in,
  right  = 1.65in,
  top    = 1.0in,
  bottom = 1.0in,
  ]{geometry}

\definecolor{shadecolor}{gray}{0.9}

\usepackage[final,expansion=alltext]{microtype}
\usepackage[english]{babel}
\usepackage[parfill]{parskip}
\usepackage{afterpage}
\usepackage{framed}

{\endMakeFramed}

\DeclareRobustCommand{\parhead}[1]{\textbf{#1}~}

\usepackage{lineno}

\usepackage{ragged2e}

\newcounter{parcount}

\usepackage{graphicx}
\usepackage[labelfont=bf]{caption}

\usepackage{booktabs}
\usepackage{multirow}

\usepackage[algoruled]{algorithm2e}
\usepackage{listings}
\usepackage{fancyvrb}
\fvset{fontsize=\normalsize}

\usepackage{natbib}

\usepackage[acronym,nowarn]{glossaries}
\makeglossaries

\definecolor{tangerine}{rgb}{0.95, 0.52, 0.0}
\definecolor{palebrown}{rgb}{0.6, 0.46, 0.33}
\definecolor{peru}{rgb}{0.8, 0.52, 0.25}
\usepackage[colorlinks,linktoc=all]{hyperref}
\usepackage[all]{hypcap}
\hypersetup{citecolor=peru}
\hypersetup{linkcolor=peru}
\hypersetup{urlcolor=peru}

\usepackage[nameinlink]{cleveref}
\creflabelformat{equation}{#1#2#3}
\crefname{equation}{eq.}{eqs.}  
\Crefname{equation}{Eq.}{Eqs.}

\lstdefinestyle{mystyle}{
    commentstyle=\color{OliveGreen},
    keywordstyle=\color{BurntOrange},
    numberstyle=\tiny\color{black!60},
    stringstyle=\color{MidnightBlue},
    basicstyle=\ttfamily,
    breakatwhitespace=false,
    breaklines=true,
    captionpos=b,
    keepspaces=true,
    numbers=left,
    numbersep=5pt,
    showspaces=false,
    showstringspaces=false,
    showtabs=false,
    tabsize=2
}
\usepackage[colorinlistoftodos,
           prependcaption,
           textsize=small,
           backgroundcolor=yellow,
           linecolor=lightgray,
           bordercolor=lightgray]{todonotes}

\DeclareRobustCommand{\parhead}[1]{\textbf{#1}~}

\usepackage{amsthm}  

\usepackage{bm}

\usepackage[colorinlistoftodos,
           prependcaption,
           textsize=small,
           backgroundcolor=yellow,
           linecolor=lightgray,
           bordercolor=lightgray]{todonotes}

\usepackage{graphicx}
\usepackage{caption}
\usepackage{subcaption}
\usepackage{wrapfig}

\usepackage{booktabs}
\usepackage{arydshln} 
\usepackage{multirow}
\usepackage{nicematrix}

\usepackage{listings}
\usepackage{fancyvrb}
\fvset{fontsize=\normalsize}

\usepackage[nameinlink]{cleveref}
\creflabelformat{equation}{#1#2#3}
\crefname{equation}{eq.}{eqs.}  
\Crefname{equation}{Eq.}{Eqs.}

\usepackage{listings}
\lstdefinestyle{alp_style}{
    commentstyle=\color{OliveGreen},
    numberstyle=\tiny\color{black!60},
    stringstyle=\color{BrickRed},
    basicstyle=\ttfamily\scriptsize,
    breakatwhitespace=false,
    breaklines=true,
    captionpos=b,
    keepspaces=true,
    numbers=none,
    numbersep=5pt,
    showspaces=false,
    showstringspaces=false,
    showtabs=false,
    tabsize=2
}

\usepackage{capt-of}

\usepackage{lipsum}

\theoremstyle{remark}
\newtheorem*{lemma*}{Lemma}

\DeclareMathOperator*{\argmax}{arg\,max}
\DeclareMathOperator*{\argmin}{arg\,min}

\usepackage{authblk}
\usepackage{enumitem} 
\usepackage{algorithmic}
\usepackage{tabularx}
\usepackage{booktabs}
\usepackage{pifont}
\usepackage[T1]{fontenc}
\usepackage{multirow}
\usepackage{threeparttable}

\usetikzlibrary{calc}
\usepackage{tikz}
\usepackage{amsmath}
\usetikzlibrary{arrows.meta, positioning, calc, decorations.pathreplacing}
\usepackage{adjustbox}

\title{\textbf{Vendi Information Gain: An Alternative To Mutual Information For Science And Machine Learning}}

\author[1, 2]{Quan Nguyen}
\author[1, 2]{Adji Bousso Dieng}
\affil[1]{Department of Computer Science, Princeton University}
\affil[2]{\href{https://vertaix.princeton.edu/}{Vertaix}}

\makeglossary 

\begin{document}
\maketitle

\begin{abstract}
\noindent In his 1948 seminal paper \emph{A Mathematical Theory of Communication} that birthed information theory, Claude Shannon introduced mutual information (MI), which he called ``rate of transmission'', as a way to quantify information gain (IG) and defined it as the difference between the marginal and conditional entropy of a random variable. While MI has become a standard tool in science and engineering, it has several shortcomings. First, MI is often intractable---it requires a density over samples with tractable Shannon entropy---and existing techniques for approximating it often fail, especially in high dimensions. Moreover, in settings where MI is tractable, its symmetry and insensitivity to sample similarity are undesirable. In this paper, we propose the \emph{Vendi Information Gain (VIG)}, a novel alternative to MI that leverages the Vendi scores, a flexible family of similarity-based diversity metrics. We call the logarithm of the VS the \emph{Vendi entropy}\footnote{We use the phrase ``Vendi entropy'' to denote the logarithm of the Vendi score, i.e. the Renyi entropy of the eigenvalues of a similarity matrix induced by a set of samples. This is distinct from the Renyi entropy of a probability distribution over the samples.} and define VIG as the difference between the marginal and conditional Vendi entropy of a variable. Being based on the VS, VIG accounts for similarity. Furthermore, VIG generalizes MI and recovers it under the assumption that the samples are completely dissimilar. Importantly, VIG only requires samples and not a probability distribution over them. Finally, it is asymmetric, a desideratum for a good measure of IG that MI fails to meet. VIG extends information theory to settings where MI completely fails. For example, we use VIG to describe a novel, unified framework for active data acquisition, a popular paradigm of modern data-driven science. We demonstrate the advantages of VIG over MI in diverse applications, including in cognitive science to model human response times to external stimuli and in epidemiology to learn epidemic processes and identify disease hotspots in different countries via level-set estimation.\\

\end{abstract}

\section{Introduction}
Mutual information (MI) \citep{shannon1948mathematical,cover1999elements} quantifies the dependence between two random variables. Formally, it is the difference between the marginal entropy of one variable and its conditional entropy given the other. MI is symmetric, mirroring the fact that statistical dependence itself is symmetric. MI is also the standard measure of information gain (IG)—the amount by which our uncertainty about one variable decreases after observing another. It has been widely applied across domains: relating neural activity to external stimuli in cognitive science \citep{panzeri1999correlations}; feature selection \citep{peng2005feature} and clustering \citep{vinh2009information} in machine learning; multimodal image registration \citep{maes2002multimodality}; and active data acquisition \citep{mackay1992information,mackay1992evidence}.

Despite its popularity, MI has several shortcomings as a measure of information gain. 
First, because it relies on Shannon entropy, it accounts for the likelihood of different outcomes but not their pairwise similarity~\citep{leinster2012measuring}. Second, when the distributions of the random variables of interest are not available in closed form, estimating MI from samples becomes increasingly difficult, especially in high dimensions and under limited-sample settings~\citep{quian2009extracting,macke2011biased,dorval2011estimating,rhee2012application,kinney2014equitability}. It has been shown that frequently employed estimators of MI can yield arbitrarily poor approximations~\citep{paninski2003estimation,molter2020limitations,song2020understanding}. Third, MI’s inherent symmetry is at odds with the directional nature of information gain~\citep{amblard2011directed,carr2019use,karmon2016biological, schroeder2004alternative}.

\begin{table}[t]
\centering
\caption{
Quantification of information gain about a channel's input upon observing its output.
\textbf{First column}: VIG reduces to MI under transmitted symbols that are dissimilar from one another.
\textbf{Second and third column}: MI considers the middle channel with three outputs to have higher capacity, as each output corresponds to fewer input values; VIG on the other hand prioritizes the last channel, agreeing with the mean absolute error when predicting the transmitted input.
}
\label{tab:channel}
\begin{tabular}{cccc}
\toprule
& \begin{tikzpicture}
    
    
    

    \node (x) at (0,2) {$x$};
    \node (y) at (0,0) {$y$};
    \node (z) at (0,-2) {$z$};
    
    \node (a) at (3,1) {$a$};
    \node (b) at (3,-1) {$b$};
    
    \foreach \a in {a, b} {
        \foreach \x in {x, y, z} {
            \draw[->] (\x) -- (\a);
        }
    }
    

\end{tikzpicture} & \begin{tikzpicture}
    
    
    
    \node (a) at (3,1) {$a$};
    \node (b) at (3,0) {$b$};
    \node (c) at (3,-1) {$c$};
    
    \node (x1) at (0,2) {$0$};
    \node (x2) at (0,1.2) {$0.1$};
    \node (x3) at (0,0.4) {$0.2$};
    \node (x4) at (0,-0.4) {$0.3$};
    \node (x5) at (0,-1.2) {$0.4$};
    \node (x6) at (0,-2) {$0.5$};
    
    \draw[->] (x1) -- (a);
    \draw[->] (x4) -- (a);
    \draw[->] (x2) -- (b);
    \draw[->] (x5) -- (b);
    \draw[->] (x3) -- (c);
    \draw[->] (x6) -- (c);
    

\end{tikzpicture} & \begin{tikzpicture}
    
    

    \node (a) at (3,1) {$a$};
    \node (b) at (3,-1) {$b$};
    
    \node (x1) at (0,2) {$0$};
    \node (x2) at (0,1.2) {$0.1$};
    \node (x3) at (0,0.4) {$0.2$};
    \node (x4) at (0,-0.4) {$0.3$};
    \node (x5) at (0,-1.2) {$0.4$};
    \node (x6) at (0,-2) {$0.5$};
    
    \draw[->] (x1) -- (a);
    \draw[->] (x2) -- (a);
    \draw[->] (x3) -- (a);
    \draw[->] (x4) -- (b);
    \draw[->] (x5) -- (b);
    \draw[->] (x6) -- (b);
    

\end{tikzpicture} \\
\midrule
$\mathrm{MI}$ & $0.176$ & $0.477$ & $0.301$ \\
$\mathrm{VIG}$ & $0.176$ & $0.338$ & $0.665$ \\
\midrule
MAE & -- & 0.150 & 0.089 \\
\bottomrule
\end{tabular}
\end{table}

This work proposes an alternative to MI called \emph{Vendi information gain} (VIG), defined as the residual Vendi entropy of a variable of interest after observing the value of another variable. Consider the example of communication channels, the application studied in \citet{shannon1948mathematical} where MI was first proposed. Each of the three diagrams in \cref{tab:channel} visualizes a communication channel, which takes in one of the elements on the left as input and returns one of the elements on the right. Each arrow indicates there is a positive probability the element at one endpoint is returned as the element at the other end point when transmitted by the channel. We assume all arrows going from the same element share the same probability, and all elements on the left are transmitted equally often by a channel. We are interested in quantifying IG about the channel's input upon observing its output. 

To compute VIG, we subtract from the initial Vendi entropy of the channel's input, the \emph{conditional} Vendi entropy for each corresponding output. This difference in Vendi entropy quantifies how much we have learned about the input upon observing the output. Crucially, VIG uses a kernel that reflects the fact that not all pairs of inputs are equally dissimilar. In this example, the second channel should yield a lower IG than the third, as estimating the input from an output observation yields a higher error in the second than in the third. This is not obeyed by MI, which treats all inputs as completely distinct, which is only true for the first communication channel.
Across the settings, VIG gives a more faithful measure of IG.

In Section \ref{sec:results}, we demonstrate that VIG overcomes MI’s limitations and outperforms it in diverse settings. We also introduce in Section \ref{sec:results} a novel VIG‐based active data acquisition framework, delivering the first fully information‐theoretic method for level‐set estimation. Finally, we derive VIG’s key properties in Section \ref{sec:theory}.



\section{Background}
The Vendi Score (VS) \citep{friedman2023vendi,pasarkar2024cousins} operates on a finite set of points $D = \{ \theta_i \}_{i = 1}^n$ sampled from domain $\Theta$.
Assuming access to a positive semidefinite kernel function $k: \Theta \times \Theta \rightarrow \mathbb{R}$, where $k(\theta, \theta) = \nolinebreak 1, \forall \theta \in \Theta$, we first compute the kernel matrix $K \in \mathbb{R}^{n \times n}$, where each entry $K_{i, j} = k(\theta_i, \theta_j)$.
The VS is then defined as:
\begin{equation}
\label{eq:vs_q}
\mathrm{VS}_q (D; k) = \exp \left( \frac{1}{1 - q} \, \log \biggl( \sum_{i = 1}^n (\overline{\lambda}_i)^q \biggr) \right).
\end{equation}
where $\overline{\lambda}_1, \overline{\lambda}_2, \ldots, \overline{\lambda}_n$ are the non-zero normalized eigenvalues of $K$, and the order $q \geq 0$ is a hyperparameter.
\citet{friedman2023vendi} discussed the various properties of the VS as a diversity metric.
For example, the VS is minimally $1$ when all samples are identical, i.e. $k(\theta_i, \theta_j) = 1, \forall i \neq j$, and achieves its maximum $n$ when all the samples are completely dissimilar, i.e. $k(\theta_i, \theta_j) = 0, \forall i \neq j$ \citep{friedman2023vendi}. When $q \in \{ 1, \infty \}$, the $\mathrm{VS}$ is defined as the limit of \cref{eq:vs_q} as $q$ approaches the value in question. \citet{pasarkar2024cousins} showed that the order $q$ controls the sensitivity of the VS to rarity: low values of $q$ result in scores that are more sensitive to rare features of the items, while high values of $q$ prioritize common features. 

The formulation of the VS above is done directly on a set of data points (e.g., a set of samples drawn from a distribution).
\citet{friedman2023vendi} further provided an alternative form named the \emph{probability-weighted} VS where, in addition to a set of samples $\{ \theta_i \}$, we are provided with the samples' corresponding probabilities, concatenated into the vector $\boldsymbol{p}$.
The kernel matrix $K$ in \cref{eq:vs_q} is replaced by a probability-weighted kernel matrix:
\begin{equation}
\label{eq:prob_weight_K}
K_{\boldsymbol{p}} = \mathrm{diag} (\sqrt{\boldsymbol{p}}) ~ K ~ \mathrm{diag} (\sqrt{\boldsymbol{p}}),
\end{equation}
from which the normalized eigenvalues $\overline{\lambda}_i$ are computed.
This probability-weighted form is useful when the exact distribution of the samples is known, as it gives the limit of the VS computed from increasing numbers of samples.

\section{Method}

We now define VIG and present its properties as an IG metric.

\subsection{Defining the Vendi Information Gain}

We call the logarithm of the VS, that is, the R\'enyi entropy of the normalized eigenvalues of the kernel matrix (or the probability-weighted kernel matrix), the  \emph{Vendi entropy}:
\begin{equation}
\label{eq:von_neumann_entropy}
H_V(D; q) = \frac{1}{1 - q} \, \log \biggl( \sum_{i = 1}^n (\overline{\lambda}_i)^q \biggr),
\end{equation}
where $q$ is the order and $D = \{ \theta_i \}_{i = 1}^n$ is a collection of samples drawn from the distribution of $\theta$ (or the values within the support of the distribution and their corresponding probabilities for the probability-weighted form). We have made the dependence on a kernel $k$ implicit. We then define VIG as the expected reduction in Vendi entropy of a random variable of interest $\theta$ after the value of another random variable $y$ is observed:
\begin{equation}
\label{eq:vig}
\mathrm{VIG}(\theta, y; q) = H_V(D; q) - \mathbb{E}_y[H_V(D_y; q)],
\end{equation}
where $D$ is a collection of samples drawn from $p(\theta)$, and $D_y$ contains samples from the conditional distribution $p(\theta \mid y)$ for a given value of $y$. VIG is directly defined on the samples; it doesn't require specifying a tractable probability distribution over them, and thus is more flexible than MI, especially in limited-data settings or high dimensions, as we'll demonstrate in our experiments.


\begin{figure*}[t]
\centering
\includegraphics[width=\linewidth]{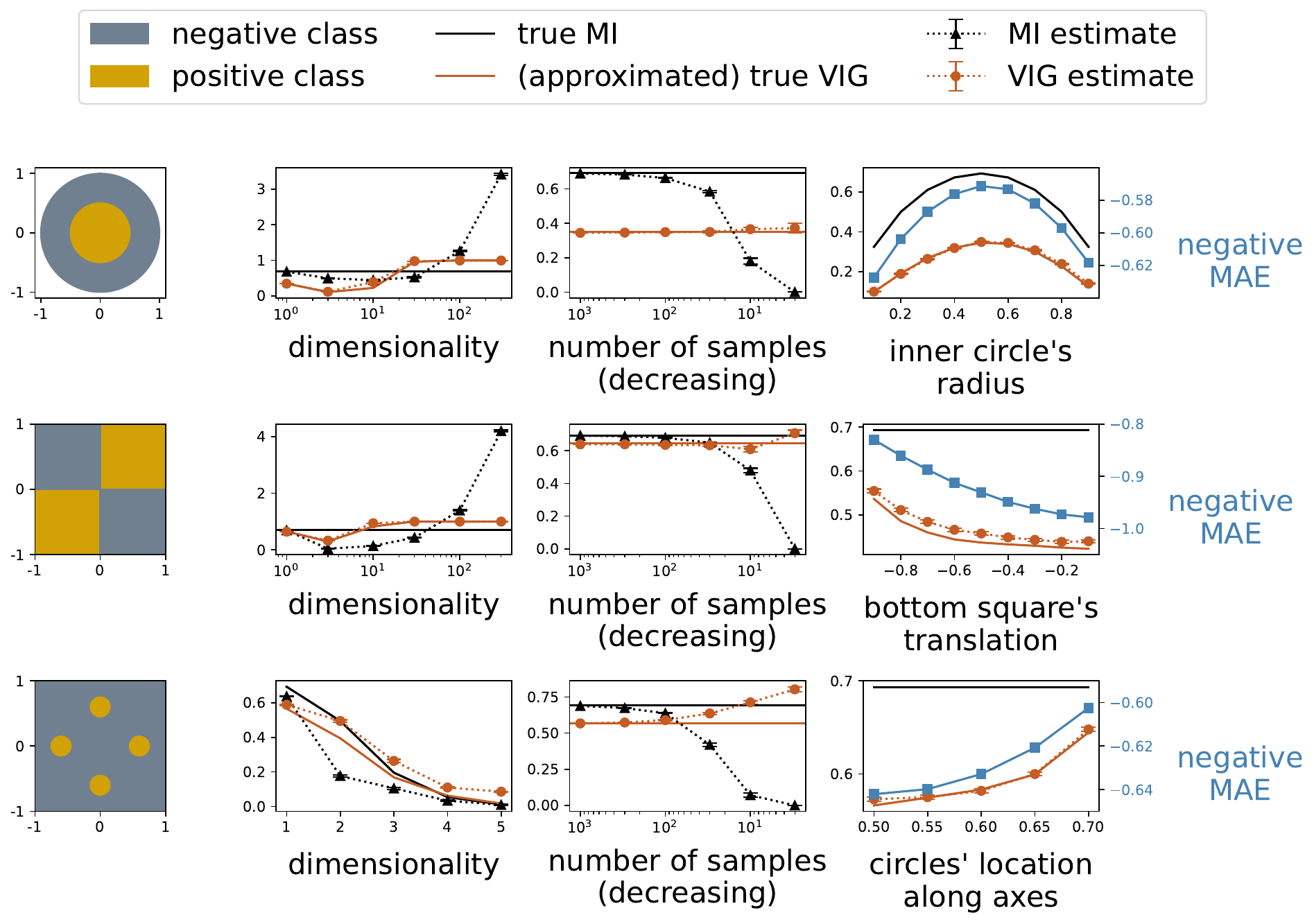}
\caption{
Illustration of MI's failure modes and VIG's benefits.
MI estimates tend to become biased in unpredictable ways as the dimensionality increases in the \textbf{second column} and degenerate to $0$ with fewer samples in the \textbf{third column}, while VIG estimates are relatively stable in both.
\textbf{Last column}: Compared to MI, VIG better corresponds to the trend of the average predictive error conditioned on the label.
}
\label{fig:compare}
\end{figure*}

We visualize MI's shortcomings in \cref{fig:compare} using three settings described by different distributions depicted in the first column.
In each case, random variable $\theta$ of interest follows a uniform distribution and has the label $y$ denoting the region it lies in.
We are interested in quantifying IG about $\theta$ by observing $y$.

The second and third columns of \cref{fig:compare} show how well estimates of MI and VIG approximate the true values under increasing dimensionality and decreasing sample size. The estimated MI is a poor approximation, becoming biased in unpredictable ways. This coincides with previous works which have found that estimating MI is a challenging endeavor in high dimensions~\citep{quian2009extracting}, with limited samples~\citep{molter2020limitations}, or under non-Gaussian, multimodal distributions~\citep{peyrard2025meta}.
On the other hand, the estimated VIG is quite stable under increasing dimensionality as well as limited samples.
In particular, VIG does not degenerate to $0$ with decreasing samples like MI.

Even when available in closed form, MI is insensitive to sample similarity and may mischaracterize IG, as demonstrated by the last column of \cref{fig:compare} where we vary the characteristics of the distributions and observe the resulting behaviors.
To approximate the behavior of a good quantification of IG, we compute the negative mean absolute error (MAE) when predicting $\theta$ conditioned on $y$.
We expect the trend in IG to correspond to MAE: IG should be high when MAE is low and vice versa.
This is not the case for MI, which does not correspond to and even conflict with MAE,
a directly result of MI's failure to account for distances that has been discussed in previous works~\citep{petty2018some,schroeder2004alternative,oliver2022information,cheng2024measuring}.
Meanwhile, VIG faithfully corresponds to the behavior of the MAE, thanks to its ability to account for similarity between samples.

\subsection{Understanding the Vendi Information Gain}
We discuss several interesting properties of VIG, contrasting them with the properties of MI when applicable.

\parhead{Reducing to $0$ with independent variables.}
From \cref{eq:vig}, we see that VIG equals $0$ when $\theta$ does not depend on $y$, as conditioning on a value of $y$ does not change the posterior samples of $\theta$.
While not explored in this work, one may reasonably use VIG as an alternative to MI to measure dependence between random variables.

\parhead{Generalization of MI.} VIG reduces to MI under completely dissimilar samples (when there is no similarity between different realizations of the variable of interest).
To see this, note that the kernel matrix $K$ is the identity matrix in such a case, and the probability-weighted kernel matrix in \cref{eq:prob_weight_K} is a diagonal matrix whose entries are the corresponding probabilities of the samples:
\begin{equation}
K_{\boldsymbol{p}} = \mathrm{diag}(\boldsymbol{p}).
\end{equation}
The eigenvalues of this kernel matrix $K_{\boldsymbol{p}}$ are the probabilities in $\boldsymbol{p}$, and Vendi entropy in \cref{eq:vs_q} becomes the R\'enyi entropy of the variable's distribution.
Under order $q = 1$, this entropy is the Shannon entropy $H$, and VIG recovers MI:
\begin{equation}
\mathrm{VIG}(\theta, y; q = 1) = H(\theta) - H(\theta \mid y) = I(\theta; y).
\end{equation}
MI is thus a special realization of VIG under distinct samples.
As shown throughout this work, this assumption of distinct samples is not applicable in many realistic scenarios, which induces undesirable behavior from MI.
VIG accounts for these scenarios where distances between samples are meaningfully defined via its kernel, and ultimately provides a richer quantification of IG.

\parhead{Boundedness.} Similar to MI, VIG ranges between $0$ minimally (when $\theta$ does not depend on $y$, and $D \equiv D_y, \forall y$) and the initial entropy maximally (when $y$ completely determines $\theta$, and $D_y$ contains one member).
However, as shown in \cref{fig:compare}, VIG will give rise to different behaviors than those from MI in other cases. 

\parhead{Asymmetry.} MI is symmetric with respect to its two inputs $\theta$ and $y$, which often yields the interpretation that, the amount of information gained about $\theta$ upon observing $y$ is equal to the amount of information gained about $y$ upon observing $\theta$.
However, this symmetry obfuscates the fundamental differences between entropy of $y$ and entropy of $\theta$, which may make reasoning about IG about these variables more challenging.
Consider a simple scenario where:
\begin{equation}
\theta \sim U[-1, 1]; \quad y = \mathbb{I} \big[ | \theta | < 0.5 \big].
\end{equation}
Here, the mutual information $I(\theta; y) = \log 2$.
In terms of IG about $y$, observing $\theta$ yields perfect information about $y$, reducing its initial discrete entropy from $H(y) = \log 2$ to $H(y \mid \theta) = 0$.
Due to MI's symmetry, we have the same reduction in the differential entropy of $\theta$: $H(\theta) = \log 2$ and $H(\theta \mid y) = 0$.
However, the same interpretation of IG does not apply anymore, as $H(\theta \mid y) = 0$ does not suggest perfect information about $\theta$ upon observing $y$.
Note that we do not argue against symmetry when MI is used as a metric to assess the dependence between two variables (in which case symmetry is indeed desirable).
This is not the same for IG, however, where we may want to model asymmetric information flow.
For instance, asymmetry has also been found to be a desirable property of IG under the context of causal inference~\citep{ay2008information,peters2017elements,simoes2024fundamental}, where there is an inherent asymmetry in a causal relationship between random variables, or in cryptography where perfect correlation that cannot be uncovered in polynomial time is not considered usable information~\citep{xu2020theory}.

Unlike MI, VIG is not symmetric: IG about $\theta$ upon observing $y$ does not necessarily equal IG about $y$ upon observing $\theta$.
This is because each of the VIG quantities depends on the structure of the distribution of the variable of interest (made explicit by the corresponding kernel functions).
Symmetry is achieved when both kernels are chosen to be the identity kernel, reducing VIG to MI; however, this choice of kernel may not be appropriate in many scenarios where there is a meaningful similarity metric between samples of $\theta$ or of $y$.
Consider the second and third communication channels of \cref{tab:channel}, where the channels' inputs $\theta$ are real-valued numbers and the outputs $y$ are distinct symbols.
As $\theta$ and $y$ are different random variables, we view it reasonable to not expect IG about $\theta$ to be equal to IG about $y$.

MI's symmetry is often used to simplify its calculations when computing $H(\theta)$ and $H(\theta \mid y)$ is challenging, such as when $\theta$ is high-dimensional and follows a multimodal distribution but $y$ has a closed-form one-dimensional distribution, so MI can be computed via $H(y)$ and $H(y \mid \theta)$.
However, its lack of symmetry does not mean that VIG will run into the same problems as MI, since VIG can be computed directly on samples of $\theta$.

\parhead{Additivity under independence.} Similar to MI, VIG is additive under independent variables.
Indeed, if $x$ and $y$ are conditionally independent from one another:
\begin{equation}
\mathrm{VIG} \big( (x, y), z \big) = \mathrm{VIG}(x, z) + \mathrm{VIG}(y, z),
\end{equation}
where $z$ is a third, observable random variable.
We prove this by showing that the Vendi entropy can be decomposed into the sum of individual terms:
\begin{equation}
H_V(D_{x, y}; q) = H_V(D_x; q) + H_V(D_y; q),
\end{equation}
where $D_x$ and $D_y$ of sizes $n$ and $m$ respectively are data sets of i.i.d.\ samples of $x$ and $y$, and $D_{x, y}$ is sampled from the joint distribution.
We note that as $x$ and $y$ are independent, the covariance matrix $K_{x, y}$ corresponding to the joint data set $D_{x, y}$ is the Kronecker product of $K_x$ and $K_y$ corresponding to individual data sets $D_x$ and $D_y$ respectively:
\begin{equation}
K_{x, y} = K_x \otimes K_y.
\end{equation}
\citet{friedman2023vendi} considered an alternative way to compute the VS, which is to first normalize the kernel matrix by its size to obtain unit trace.
This normalized matrix yields eigenvalues that are the same normalized eigenvalues in \cref{eq:vs_q} which sum to $1$.
We now consider the eigenvalues of $\frac{1}{nm} K_{x, y}$ denoted by $\{ {\lambda_x}_i {\lambda_y}_j \}_{1 \leq i \leq n, 1 \leq j \leq m}$, where each ${\lambda_x}_i$ (resp.\ ${\lambda_y}_j$) is an eigenvalue of the covariance matrix $\frac{1}{n} K_x$ (resp.\ $\frac{1}{m} K_y$).
We observe that these eigenvalues also sum to $1$:
\begin{equation}
\sum_{1 \leq i \leq n, 1 \leq j \leq m} {\lambda_x}_i {\lambda_y}_j
= \sum_{1 \leq j \leq m} \left( {\lambda_y}_j \sum_{1 \leq i \leq n} {\lambda_x}_i \right)
= \sum_{1 \leq j \leq m} {\lambda_y}_j (1)
= 1,
\end{equation}
and thus the Vendi entropy of $D_{x, y}$ may be computed as:
\begin{equation}
\begin{split}
H_V(D_{x, y}; q)
& = \frac{1}{1 - q} \, \log \biggl( \sum_{1 \leq i \leq n, 1 \leq j \leq m} ({\lambda_x}_i {\lambda_y}_j)^q \biggr) \\
& = \frac{1}{1 - q} \, \log \left( \left( \sum_{1 \leq i \leq n} ({\lambda_x}_i)^q \right) \left( \sum_{1 \leq j \leq m} ({\lambda_y}_j)^q \right) \right) \\
& = \frac{1}{1 - q} \, \log \left( \sum_{1 \leq i \leq n} ({\lambda_x}_i)^q \right) + \frac{1}{1 - q} \, \log \left( \sum_{1 \leq j \leq m} ({\lambda_y}_j)^q \right) \\
& = H_V(D_x; q) + H_V(D_y; q).
\end{split}
\end{equation}
VIG's additivity is valuable as decomposing the joint covariance matrix $K_{x, y}$ is more expensive than decomposing the individual matrices $K_x$ and $K_y$.

\parhead{Dependence on variable's spread.} As illustrated in \cref{fig:compare}, VIG, unlike MI, is a function of the \emph{values} within the support of the random variable of interest, or more specifically, the distance between one value to another.
This dependence on the spread of the variable's distribution leads to the main difference in behavior between VIG and MI, which is at its starkest when the distribution becomes multimodal.

\section{Experiments}
\label{sec:results}

We now demonstrate VIG's superiority over MI in several applications. In particular, we use VIG to develop a unified framework for active data acquisition, including a novel, first-ever information-theoretic approach to level-set estimation. 

\subsection{Modeling human response times to external stimuli}


\begin{figure}[!ht]
\centering
\includegraphics[width=\linewidth]{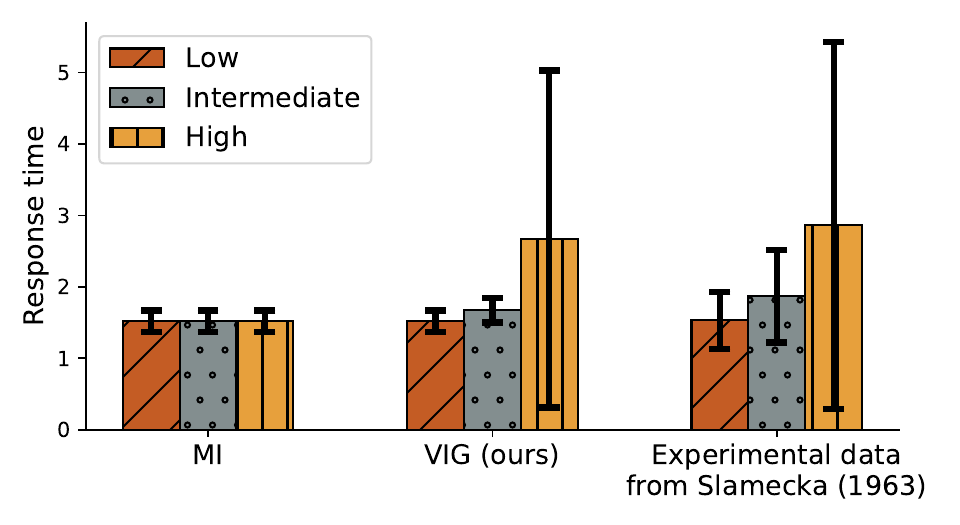}
\caption{
Simulated response times as a function of similarity between possible messages by MI and VIG against observed experimental data from \citet{slamecka1963choice}.
MI's insensitivity to inter-message similarity prevents the simulation of different behaviors under different conditions, while VIG's simulations closely match observed real-world data.
}
\label{fig:neuro_compare}
\end{figure}

\citet{christie2023information} proposed an information-theoretic framework to model the time it takes to respond to external stimuli, a central quantity in many important concepts in behavioral psychology such as priming effects~\citep{hart2010emotional} and the working memory capacity~\citep{draheim2016combining}.
The authors developed a simple model of neural decoding in which external stimuli---the intended stimulus and other sources of noise---are observed as messages generated by Poisson processes.
Denote the set of all sources of stimuli as $M$; the observer maintains a Bayesian belief about which message $m \in M$ corresponds to the true stimulus they should respond to, which is updated over time based on observed data.

While messages come in, IG about the true stimulus is measured by MI between the intended message and the current data $I(m, D_t)$, where $D_t$ denotes the observed data at time $t$.
Once MI exceeds a certain threshold $h$, the decoding process is considered complete, and the observer returns the \emph{maximum a posteriori} (MAP) estimate of the intended message.
The time taken for MI to reach the threshold and return the MAP estimate is considered the response time $\mathrm{RT}$,
which \citet{christie2023information} showed models many important phenomena such as the Hick--Hyman Law~\citep{hyman1953stimulus}, the Power Law of Practice~\citep{dayan2000learning}, and the speed--accuracy tradeoff~\citep{heitz2014speed}.

However, this model has its own limitations.
As MI treats messages as completely distinct, it ignores the similarity between any two given messages, which rules out scenarios in which similarity between responses is meaningful.
Consider the concept of \emph{confusability} in psychology, which dictates that the time taken to make a choice between alternatives increases with the similarity between those alternatives.
For instance, \citet{dember1957relation} and \citet{slamecka1963choice} demonstrated a negative correlation between response time in picking out words and perceived distance between those words, measured by dissimilarity in word-meaning, while \citet{podgorny1979reaction} found a similar relationship between response time and visual similarity between letters of the alphabet.
Being able to model inter-message similarity is therefore crucial in developing a high-fidelity model of human response time.

We propose using VIG as the criterion with which we decide when the decoding process in the model of \citet{christie2023information} is complete.
The modification allows us to account for similarity between different possible messages by constructing the kernel matrix for the possible messages.
This decoding rule can then be plugged into the model by \citet{christie2023information} to replace MI and induce a different behavior in the resulting model.

We demonstrate VIG's ability to model the concept of confusability by performing the simulation with the model by \citet{christie2023information}, transmitting three symbols $A$, $B$, and $C$ with varying degrees of similarity.
In particular, symbol $C$ is modeled as being distinct from $A$ and $B$, while the similarity between $A$ and $B$ varies from low to intermediate to high.
We repeat the same simulation 100 times and show in \cref{fig:neuro_compare} the distribution of simulated response times resulting from using MI and VIG, as well as the real-world data reported by \citet{slamecka1963choice} from their word-meaning experiments as a reference.
From the real-world experimental data, we observe the natural trend that the higher the similarity between the stimuli, the harder it is to distinguish between them, and thus the longer the response time is.
This trend cannot be replicated using MI but is faithfully simulated by VIG, demonstrating VIG's advantage.

\subsection{Vendi information gain leads to a general unified framework for active data acquisition}

Active data acquisition is a long-standing problem spanning diverse fields such as psychology \citep{myung2013tutorial}, diagnostic audiometry \citep{gardner2015bayesian}, physics \citep{dushenko2020sequential}, and materials science \citep{novick2024probabilistic}.
Taking on different names in different communities such as Bayesian experimental design~\citep{lindley1956measure}, design of experiments~\citep{pukelsheim2006optimal}, or active learning~\citep{settles2009active}, active data acquisition denotes the task of actively choosing data to learn about some underlying process.
Given the random variable $\theta$ representing a quantity we wish to learn about (e.g., the unknown transmission rate of a new disease, the location of a hidden energy source, the regions in which air pollution exceeds a threshold), we construct a prior $p(\theta)$ and an observation model that yields a likelihood $p(y \mid x, \theta)$ for possible labels $y$ given the data point $x$ we control.
We then obtain new data in a sequential manner---accumulating observations $\mathcal{D} = \{ (x_i, y_i) \}_{i = 1}^n$ in the process---to learn as much about $\theta$ as possible.

\begin{figure*}
\centering
\begin{adjustbox}{width=\textwidth}
\input{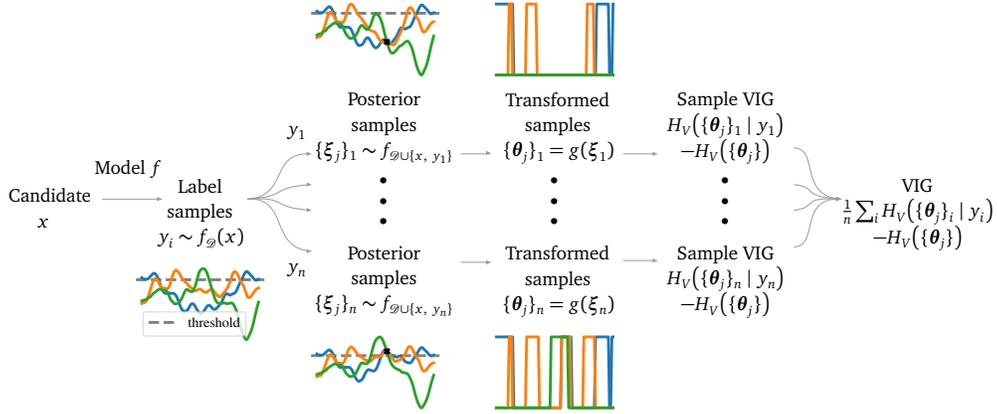}
\end{adjustbox}
\caption{
Illustration of Vendi information gain for active data acquisition.
For each candidate $x$ we may query, we first draw Monte Carlo samples of its label $\{ y_i \}$ according to our predictive model trained on observed data $f_\mathcal{D}$.
Conditioned on each sample, we then draw fantasized samples of all labels within the search space $\boldsymbol{\xi}_i$, which are transformed by function $g(\cdot)$ to yield samples of the quantity of interest $\boldsymbol{\theta}_i$.
Information gain is then computed on each set of samples, and the average across different fantasized labels gives the expected VIG.
}
\label{fig:diagram}
\end{figure*}

Traditionally, active data acquisition defines the utility of a collected data set $\mathcal{D}$ as the negative posterior entropy $-H(\theta \mid \mathcal{D})$, expressing preference for having more knowledge (and thus lower entropy) about $\theta$.
This is equivalent to choosing the data point that possesses the largest MI with $\theta$, as MI is the difference between the posterior and the fixed prior entropy.
In many cases, the posterior distribution $p(\theta \mid \mathcal{D})$ is not available in closed form, and we are required to approximate the posterior entropy $H(\theta \mid x, y)$ using samples.
Here, estimating $H$ given limited samples is a challenging task, especially if $\theta$ is high-dimensional, and previous works have been limited to simple models such as linear models~\citep{lindley1956measure,mackay1992information} to simplify the calculation of $H$.

We propose using VIG to quantify IG instead, seeking the data point that lowers the posterior Vendi entropy for $\theta$ the most. 
VIG is computed with samples of $\theta$ without requiring further assumptions about its distribution.
This gives rise to a flexible framework for active data acquisition in which we draw posterior samples from a predictive model trained on observed data, apply the necessary transformations to obtain samples of $\theta$, and compute IG based on these samples alone.
This procedure is visualized in \cref{fig:diagram}, where, as an example, $\theta$ is the binary output of a threshold transformation on a continuous function; again, VIG makes no assumption about the structure of the distribution of $\theta$.

\begin{figure*}[t]
\centering
\includegraphics[width=\linewidth]{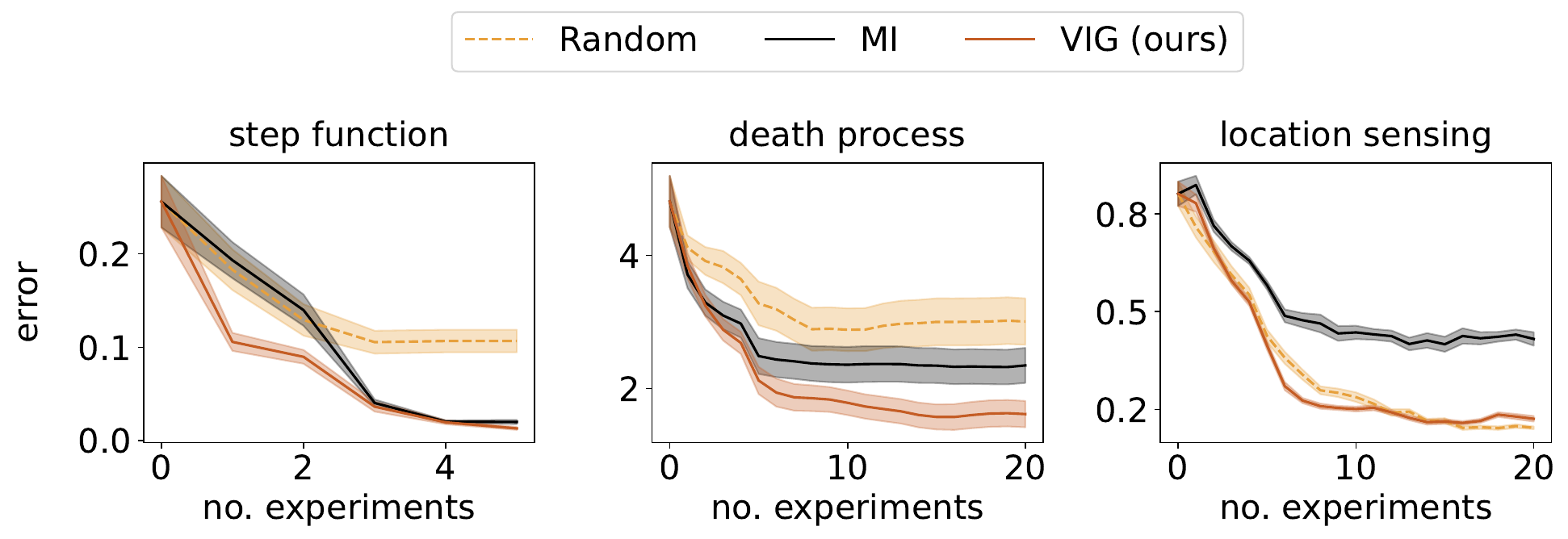}
\caption{
Average estimation error ($\pm$ 1 standard error) across 50 repeats. VIG is competitive against MI, at times outperforming MI by a large margin.
}
\label{fig:bed_results}
\end{figure*}

We compare the performance of VIG against MI as data acquisition strategies using the three tasks: learning a step function whose threshold is unknown, learning a death process that models the rate at which healthy individuals become infected with a disease, and learning the locations of two hidden energy sources by measuring the emitted signals.
We also include random sampling as the baseline data acquisition strategy.
At each iteration of the loop, we measure the absolute error between the current \emph{maximum a posteriori} (MAP) estimate of $\theta$ and its true value.
\cref{fig:bed_results} visualizes this average error (and one standard error) across the repeats in our three tasks, where we see that VIG (under $q = 1$) achieves competitive performance against MI, sometimes outperforming MI by a large margin.
Interestingly, MI is outperformed by even random sampling in the location sensing task.
We hypothesize this is due to the complex landscape of $p(\theta)$, a 4-dimensional vector encoding the coordinates of the two energy locations: not only are the first and second pairs of coordinates interchangeable, distances between locations also play a key role in reasoning about $\theta$, exacerbating the failure mode of MI.

Level-set estimation (LSE) is a prominent instance of active data acquisition where we aim to learn whether the labels of given data points are above or below a certain threshold.
A primary application is in environmental monitoring, to identify regions with high levels of e.g. humidity, pollution, or oil spills~\citep{gotovos2013active}.
Barring any limit of the number of obtainable labels, we may simply request all possible labels to be known, fully determining the regions with data exceeding the threshold.
However, measurements are often expensive (in terms of money, time, or some other risks), and we only seek a small number of data points whose labels yield the most information about the space.

The quantity of interest $\theta$ in LSE is a binary vector that indicates whether each label in the space exceeds the threshold (\cref{eq:theta_lse}), and one may seek to take an IG approach to collect data to learn about this vector.
MI is a common tool for measuring IG but has not been successfully applied to this problem.
This is because the dimensionality of $\theta$ is equal the number of data points in the search space (each point corresponds to an element in the vector).
This dimensionality can be prohibitively high in practice (for example, it is on the order of 1,000+ in our experiments), which, as \cref{fig:compare} demonstrates, prevents MI from being approximated with high fidelity.
Overall, an IG-based approach has not been attempted in previous LSE works, and many have chosen to forgo the IG formulation altogether, and design simpler, cheaper-to-compute strategies.
This work applies an IG criterion to tackle LSE \emph{for the first time}, demonstrating VIG's flexibility as an IG metric.

We compare the performance of VIG against the following existing LSE policies.
The widely used STRADDLE policy by \citet{bryan2005active} targets the location we are the most uncertain about whether the corresponding label exceeds the threshold.
To avoid repeatedly querying a boundary region and consequently getting stuck, \citet{bryan2005active} added an exploration term to encourage inspecting regions far away from observed data.
Another established baseline, which takes on the moniker ``LSE'' to match the task, was proposed by \citet{gotovos2013active} as an alternative to STRADDLE.
This policy chooses the point that maximizes a different quantification of uncertainty in whether a label exceeds the threshold, and, with a slight modification, generalizes STRADDLE~\citep{gotovos2013active}.
Contrasting these policies against the VIG framework, the former simplify the task of jointly reducing uncertainty in the entire search space to targeting the one single element with the highest uncertainty.
This simplification yields strategies that are cheaper to compute but exhibit suboptimal behavior.
We also employ the \emph{uncertainty sampling} heuristic, a common general-purpose active learning policy that queries the data point with the highest uncertainty in the label.
Finally, we include the MI criterion approximated using samples of $\theta$ (the same number of samples used by VIG).

We conduct two sets of experiments.
In the first, synthetic functions are samples from a Gaussian process~\citep{rasmussen2006gaussian}, and the dimensionality of a problem varies as $d \in \{ 1, 2, 3, 5 \}$.
Our second set of experiments uses the disease hotspot data curated by \citet{andrade2020finding}, which contains four individual data sets describing the prevalence of schistosomiasis in C\^{o}te d'Ivoire and Malawi and of lymphatic filariasis in Haiti and the Philippines.
The objective is to identify the regions within each country whose prevalence of a disease exceeds a critical threshold.

\begin{figure*}[t]
\centering
\includegraphics[width=\linewidth]{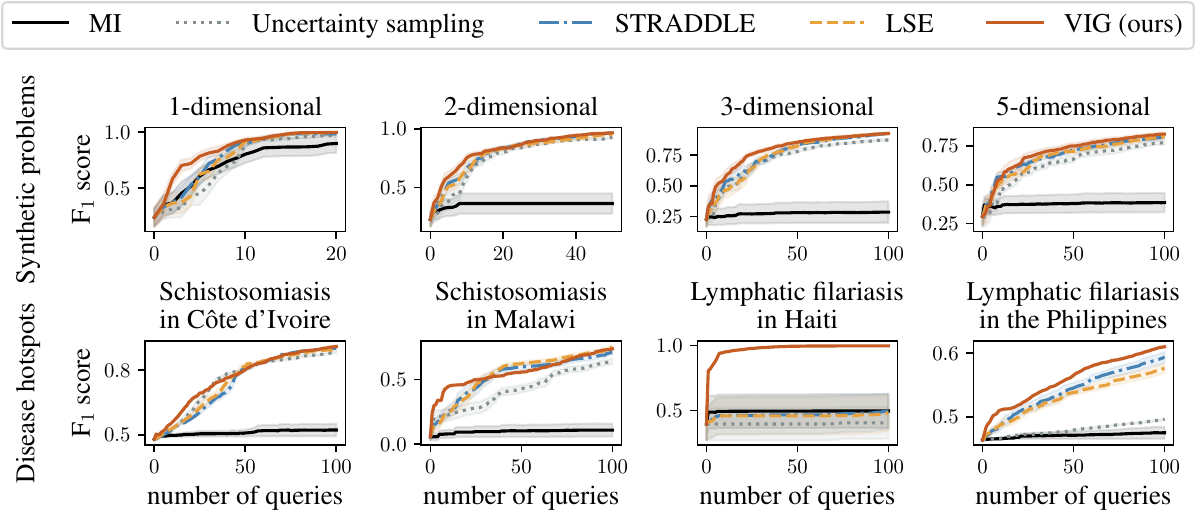}
\caption{
Average F$_1$ scores and standard errors achieved by LSE policies as a function of the number of queries.
MI performs the worst, while VIG consistently achieves the highest F$_1$ scores, sometimes significantly outperforming baselines.
}
\label{fig:lse_results}
\end{figure*}

\begin{figure*}[!ht]
\centering
\includegraphics[width=\linewidth]{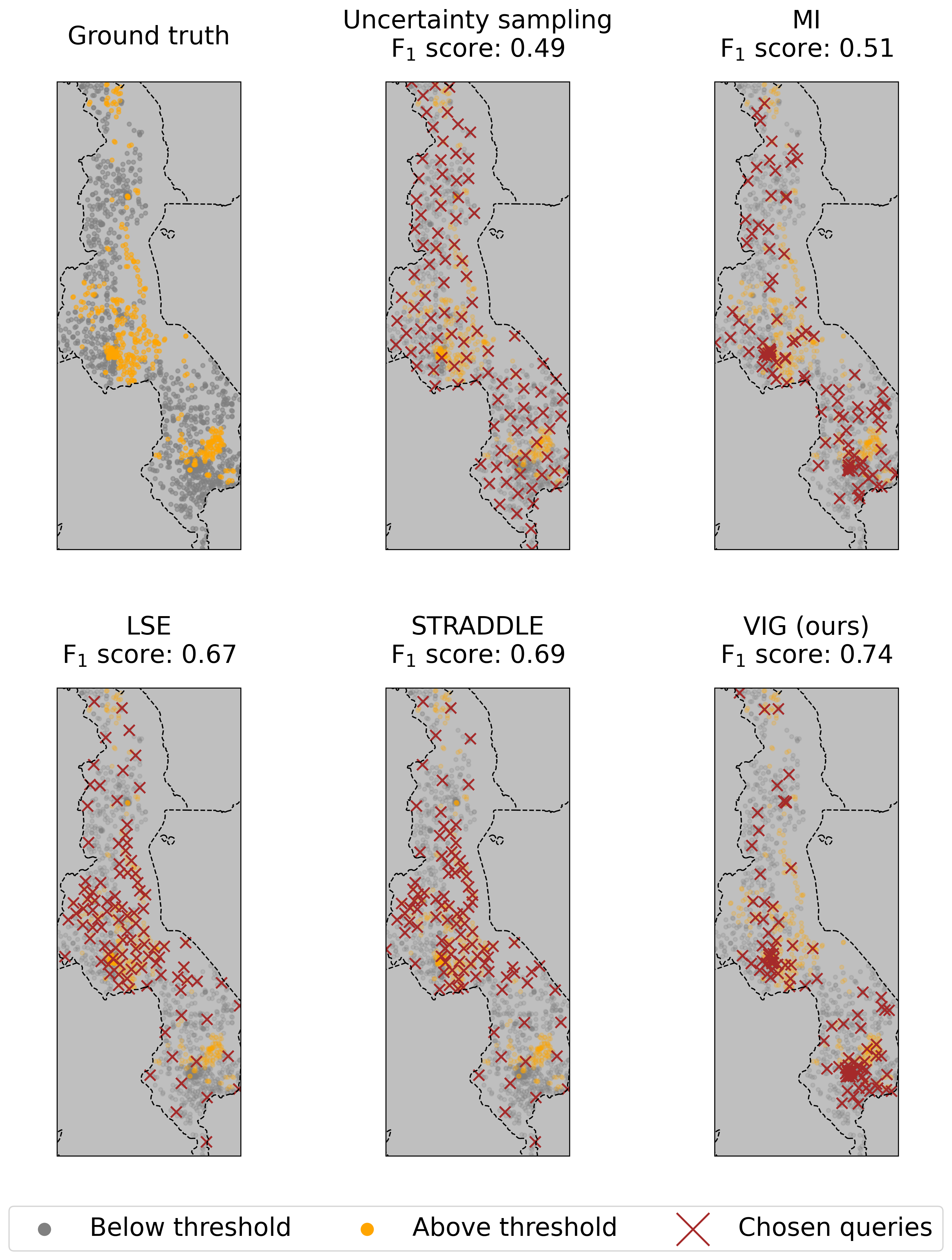}
\caption{
Behavior comparison between different policies in one illustrative run under the Schistosomiasis in Malawi experiments.
MI wastes many queries on easy-to-classify regions, while VIG focuses on multiple disease hotspots where classification is challenging, leading to a higher F$_1$ score than other baselines.
}
\label{fig:lse_runs}
\end{figure*}

As standard in the LSE literature, we record the F$_1$ score of the predictive model trained on collected data to predict whether each label exceeds the threshold.
\cref{fig:lse_runs} show this F$_1$ score by each policy under different settings, where VIG achieves a competitive performance, consistently yielding higher F$_1$ scores than the other baselines.
STRADDLE and LSE outperform uncertainty sampling as expected, as the former are specifically designed for LSE.
MI is often the worst-performing policy, which verifies the expectation from previous works that a full IG approach for LSE cannot be feasibly realized with MI.

To demonstrate the beneficial behavior of VIG, \cref{fig:lse_runs} compare the different policies under one illustrative run from the Schistosomiasis in Malawi experiment.
Here, we observe distinct behaviors: uncertainty sampling scans the space evenly to minimize the model's uncertainty, while MI wastes many queries on easy-to-classify regions.
LSE and STRADDLE allocate considerable queries to the central region, covering a large cluster of high-prevalence locations.
VIG focuses on the boundaries between low- and high-prevalence regions, targeting both the central and another large cluster in the south.
This beneficial behavior is a result of VIG's flexible reasoning---without any modification to the criterion to guard against undesirable behavior like in STRADDLE and LSE---and ultimately leads to better predictions.

Finally, we perform ablation studies on the effect of the order $q$ has on VIG in \cref{fig:bed_q,fig:lse_q} and find that performance remains fairly robust across a wide range of values for $q$, suggesting that VIG's main advantage over other policies stems from its IG reasoning for active data acquisition tasks.

\section{Related Work}
\label{sec:related}

VIG is based on the Vendi score, provides an alternative to MI for measuring IG, and leads to a unified framework for active data acquisition.
We highlight how VIG differs from methods in each of those areas.

\parhead{Mutual information.} Previous works have investigated the failure modes of MI. Indeed, researchers have highlighted the challenge in accurately estimating MI in an unbiased way, especially in high dimensions~\citep{macke2011biased,dorval2011estimating,rhee2012application,kinney2014equitability,cheng2024measuring}.
Many studies have also found MI estimators to take on undesirable pathological behaviors under multimodal, non-Gaussian distributions~\citep{czyz2023beyond,peyrard2025meta}, and even the identification of these undesirable behaviors of MI is still an active area of search.
Approximating MI may require exponentially many samples~\citep{quian2009extracting}, and even when it can be approximated with high fidelity or computed exactly, MI's insensitivity to random variables' spread may mischaracterize IG~\citep{petty2018some,oliver2022information,cheng2024measuring}.
We echo the findings by \citet{schroeder2004alternative}, who argued that MI is a good IG metric for communication systems when there is no need to model similarity between transmitted symbols, but fails in other contexts.

Various works have considered alternatives to MI that address its drawbacks.
However, these proposals are tailored to specific applications thus do not generalize to a variety of tasks where MI is employed.
For instance, \citet{torkkola2003feature} considered a version of MI that accounts for the similarity between pairs of samples (something VIG inherently does) using a Parzen estimator as part of a feature extraction procedure.
For image registration, \citet{haber2005beyond} calculated a distance-aware MI-based quantity and showed that this alternative outperforms MI under multi-modal data.
In hierarchical clustering where MI has been found to inappropriately favor large clusters under high dimensions, \citet{marrelec2015bayesian} used the empirical covariance matrix, like VIG, to correct the cluster-favoring behavior, while \citet{hafemann2024novel} addressed MI's insensitivity to spread by incorporating each data point's distances to its nearest neighbors when analyzing LiDAR sensors.
\citet{meister2024towards} computed a similarity-weighted version of Shannon entropy to model contextual word predictability.
Similar to findings in our response time application, modeling similarity between words better bridges the gap between information theoretic simulations and real-world data from cognitive science compared to MI.
\citet{liu2022fair} employed a distance-aware measure that also uses the covariance matrix of the samples to sidestep MI's intractability for learning fair representations.
\citet{kala2021global} pointed out that differential entropy may take on negative values, resulting in unintuitive IG values for sensitivity analysis tasks under MI, and \citet{petty2018some} made similar observations in the context of modeling atmospheric variables.
Vendi entropy is strictly positive, and thus VIG does not share this problem.

\parhead{Alternative information measures.} 
Previous works have pointed out the deficiencies of MI as an information measure and proposed other measures.
Of note is \citet{xu2020theory}, who presented a family of $\nu$-information measures that generalizes MI and accounts for computational constraints in modeling the relationship between the latent and the observable variables, a direction orthogonal to this work.
Their $\nu$-information measures are concerned with the observable variable's value in predicting the latent variable, which is related to the goal of minimizing predictive error when using IG as a criterion to select data.
$\nu$-information, like MI, requires a tractable density over the samples and does not account for sample similarity.

A related concept is the coefficient of determinations ($R^2$), which \citet{xu2020theory} showed $\nu$-information generalizes.
$R^2$ measures the difference between the initial variance and the residual of a model's predictions, spiritually similar to VIG measuring the difference between the initial and posterior distance-aware Vendi entropy.
Variance alone, though, is inadequate in measuring information, especially under non-Gaussian, multimodal distributions.

\citet{degroot1962uncertainty} pointed out that MI is not the only information measure for active data acquisition problems, and derived a framework for choosing an information measure based on a prediction loss.
However, this procedure is quite involved and needs to be implemented for each data-acquisition instance, while VIG offers a unified framework that generalizes across different problems.

\parhead{Information gain for active data acquisition.}
IG offers a natural way to guide active data acquisition to maximize knowledge about a quantity of interest~\citep{lindley1956measure,mackay1992information,mackay1992evidence}.
IG-based approaches are to be contrasted against a popular learning policy called ``Bayesian active learning by disagreement'', or BALD~\citep{houlsby2011bayesian}, which minimizes uncertainty in the model's parameters and may lead to suboptimal decisions in various scenarios~\citep{smith2023prediction}.
While IG has been traditionally measured by MI, as we demonstrated, the characteristics of MI may lead to misquantification of IG.
Instead of the (differential) entropy, VIG uses the Vendi Score to quantify uncertainty, addressing MI's drawbacks.

Under the context of LSE, maximizing IG by MI has not been previously realized due to MI's high computational cost, and most policies are constructed from various heuristics to target regions at the boundary, and require hyperparameters to not get stuck at a particularly complicated boundary~\citep{bryan2005active,gotovos2013active}.
Various directions in LSE are orthogonal to our contributions and thus may be combined with VIG.
For example, \citet{ha2021high} used Bayesian neural networks as surrogate models for high-dimensional functions.
As these neural networks quantify uncertainty with samples produced via stochastic forward passes, they are well positioned to work with VIG, which operates entirely on samples.
Other directions may generalize VIG to \emph{implicit} LSE, where the level-set threshold is unknown and is to be learned while collecting data.

\parhead{Vendi scoring.}
The VIG criterion is based on the Vendi Score~\citep{friedman2023vendi,pasarkar2023vendi}.
Since its inception, the VS has been applied to various settings such as enhancing the diversity of generative models' output~\citep{berns2023towards}, tuning large language models~\citep{wu2023self}, encouraging exploration in molecular simulations~\citep{pasarkar2023vendi}, evaluating cultural diversity of text-to-image models beyond existing criteria such as faithfulness and aesthetics~\citep{hall2024towards,senthilkumar2024beyond},
feature ranking~\citep{mousavi2024vsi},
evaluating diversity of prompt-based generative models~\citep{jalali2024conditional},
developing novel RAG pipelines for LLMs~\citep{rezaei2025vendi},
improving the robustness of sequence generative models to varying noise levels~\citep{rezaei2025alpha}, 
and Bayesian optimization~\citep{liu2024diversity,nguyen2024quality}. Most recently, \citet{pasarkar2025vendiscope} proposed the Vendiscope, a method based on the VS to explore and analyze the composition of large-scale datasets. We leverage the VS to propose a novel IG measure that addresses many of MI's drawbacks and yields novel frameworks for active data acquisition and level-set estimation.

\section{Discussion}
\label{sec:discussion}

We introduced a new measure of information gain (IG) called \emph{Vendi Information Gain} (VIG). Unlike mutual information (MI)---the default measure of IG in the scientific literature---VIG accounts for similarity by leveraging the Vendi Score and does not require specifying a density over samples with tractable Shannon entropy; it relies solely on samples, making it more reliable and flexible than MI. We demonstrated the utility of VIG in diverse applications such as modeling human response time to external stimuli, learning epidemic processes, and detecting disease hotspots via level-set estimation (LSE), where VIG consistently outperforms MI, showcasing its benefits as a new IG metric. 

Calculating VIG requires specifying a kernel that quantifies the similarity between any two data points. The choice of this kernel fully depends on the user. For example, we choose the Gaussian kernel to model the various real-valued quantities in our experiments and a Hamming-based kernel for binary vectors in the LSE application. Previous works using the VS have examined other choices. For example, \citet{friedman2023vendi} used a cosine similarity for an image embedding obtained via a neural network and n-gram-based kernels for text applications; \citet{nguyen2024quality} considered the Tanimoto kernel for molecular fingerprints;
\citet{pasarkar2023vendi} applied various transformations to their data as input to the Gaussian kernel to ensure symmetry invariance for molecular systems;
\citet{liu2024diversity} designed a novel kernel specifically for metal-organic frameworks.

Calculating VIG involves decomposing the kernel matrix $K$ of the samples, which scales cubically with sample size. The cubic cost is justified in settings similar to LSE, where labeling data is typically the expensive bottleneck; otherwise, we may resort to column sampling methods \citep{williams2000using} to speed up computation as discussed in \citep{friedman2023vendi}. Under a dot product-type kernel function, such as cosine similarity, this cost can be reduced~\citep{friedman2023vendi}. In settings where the distribution of the variable of interest is known, the probability-weighted form of VIG in \cref{eq:prob_weight_K} may be leveraged to reduce the size of the similarity matrix.

In active data acquisition, VIG requires drawing fantasized samples of the label vector conditioned on each label sample $y_i$ of a given candidate $x$, and many of these samples from $p(y \mid x)$ are used so that the average is a good approximation of the expectation in \cref{eq:vig}. This sampling process could prove prohibitively expensive when repeated for each possible candidate: we use a Gaussian process (GP) as our predictive model in LSE, and the cost of drawing posterior samples from a GP is cubic in the size of the search space, $O(N^3)$, as each sample corresponds to a realization of the entire label vector.
We take advantage of the pathwise conditioning framework for GPs~\citep{wilson2021pathwise}, which appeals to Matheron’s update rule to derive the posterior of a GP on the level of samples instead of distributions. More concretely, we draw samples from the GP prior \emph{once}, and condition them on training and fantasized data as needed. This conditioning step scales cubically with respect to the size of the observed data, that is, $O(| \mathcal{D} |^3)$, which is significantly lower than $O(N^3)$, especially given the assumption that the labels $y$ are expensive to obtain.

\section{Code Availability}
\label{sec:code-availability}

The human response time model is made available by \citet{christie2023information}.
All other code will be made available upon publication.

\section{Data Availability}
\label{sec:data-availability}

The disease hotspot data is made available by \citet{andrade2020finding}.
Other data are simulated using code that will be made available upon publication.

\subsection*{Acknowledgements}
Adji Bousso Dieng acknowledges support from the National Science Foundation, Office of Advanced Cyberinfrastructure (OAC): \#2118201. She also acknowledges Schmidt Sciences for the AI2050 Early Career Fellowship and DataX via Princeton's Center for Statistics and Machine Learning (CSML). 

\bibliographystyle{apa}
\bibliography{arxiv}

\newpage

\appendix

\section{Details on experiments}
\label{sec:theory}

We present the mathematical models in the analyses and the implementation of VIG in the respective applications.

\subsection{Visualizing mutual information's failures}
We describe the definitions of the distributions of the random variables $\theta$ and $y$ that are visualized in two dimensions in \cref{fig:compare}.
\begin{itemize}
\item In the first row, $\theta$ is uniformly distributed within the unit sphere in $d$ dimensions.
In the second and third column, the radius of the inner sphere is computed so that the volumes of the positive and negative regions are equal: the two values of $y$ are equally likely, yielding an MI of $\mathrm{MI}(\theta; y) = \log 2$.
\item In the second row, $\theta$ is uniformly distributed within the cube $[-1, 1]^d$ and the class label $y$ is defined to be the sign of the product of the coordinates of $\theta$ in the second and third columns.
\item In the third row, $\theta$ is again uniformly distributed within the cube $[-1, 1]^d$, and there are two spheres with radius $0.2$ along each axis; these spheres are located at coordinate $0.5$ in the first column.
\end{itemize}

\subsection{Communication channels}

In the first column of \cref{tab:channel}, the transmitted data are treated as individual symbols, which we model as completely dissimilar from one another using the identify kernel $k$ that returns $1$ if the two inputs are the same element and $0$ otherwise.
In the other columns, the channel's inputs take on numerical values, and we use a simple Gaussian kernel
\begin{equation}
\label{eq:gaussian_kernel}
k(\theta_1, \theta_2) = \exp \left( - \frac{(\theta_1 - \theta_2)^2}{2} \right).
\end{equation}

\subsection{The human response time model}

In the response time model proposed by \citet{christie2023information}, IG about the true stimulus is measured by MI between the intended message and the current data $I(m, D_t)$, where $D_t$ denotes the observed data at time $t$.
Once MI exceeds a certain threshold $h$, the decoding process is considered complete, and the observer returns the \emph{maximum a posteriori} (MAP) estimate of the intended message, thus responding to the stimuli.
The time taken for MI to reach the threshold and return the MAP estimate acts as the response time $\mathrm{RT}$:
\begin{equation}
\mathrm{RT} = \min \{ t \mid I(m, D_t) > h \}.
\end{equation}
We use VIG as the criterion with which we decide the decoding process in the model of \citet{christie2023information} is complete:
\begin{equation}
\mathrm{RT} = \min \{ t \mid \mathrm{VIG}(m, D_t) > h \},
\end{equation}
where $h$ is the user-specified threshold for IG.

To gain insight into this difference, we consider a simple toy message-decoding scenario with three messages $M = \{ A, B, C \}$, where we place a uniform prior on the intended message:
\begin{equation}
p(m = A) = p(m = B) = p(m = C) = \frac{1}{3}.
\end{equation}
Here, message $A$ and $B$ are structurally similar in that 
it is harder for the observer to distinguish $A$ and $B$ than it is to distinguish $A$ and $C$ or $B$ and $C$.
We model this structure by constructing a kernel matrix expressing similarity between the different possible messages as:
\begin{equation}
K = \begin{bmatrix}
1 & 0.5 & 0 \\
0.5 & 1 & 0 \\
0 & 0 & 1
\end{bmatrix},
\end{equation}
where we set the similarity between $A$ and $B$ as $K(A, B) = 0.5$, and message $C$ is treated as completely distinct from the other two.

Suppose we aim to compare the amount of information gained under two different possible data sets of observed stimuli $D_{t_1}$ and $D_{t_2}$, where the first data set $D_{t_1}$ yields the following posterior probabilities:
\begin{equation}
p(m = A \mid D_{t_1}) = 0.8; \quad p(m = B \mid D_{t_1}) = 0.1; \quad p(m = C \mid D_{t_1}) = 0.1,
\end{equation}
and the second $D_{t_2}$ yields:
\begin{equation}
p(m = A \mid D_{t_2}) = 0.1; \quad p(m = B \mid D_{t_2}) = 0.1; \quad p(m = C \mid D_{t_2}) = 0.8,
\end{equation}
that is, $m = A$ is the most likely intended message given $D_{t_1}$, while it is $m = C$ under $D_{t_2}$.
Intuitively, determining $m = A$ as the intended message should take more effort than finding out $m = C$ is the intended message.
This is because one needs to distinguish between $A$ and $B$, which is a harder task than distinguishing between $C$ and the other messages, due the high similarity between $A$ and $B$.
Overall, $D_{t_1}$ should yield more information than $D_{t_2}$ for us to be as confident about $m = A$ as about $m = C$.

We now compute the IG yielded by these two data sets under the two measures MI and VIG.
First, we see that MI treats these two data sets as yielding the same amount of IG:
\begin{equation}
I(m, D_{t_1}) = I(m, D_{t_2}) = H(m) - H(m \mid D_{t_1}) = H(m) - H(m \mid D_{t_2}) \approx 0.663.
\end{equation}
We then compute VIG (with $q = 1$) as:
\begin{align}
\mathrm{VIG}(m, D_{t_1}; q = 1) & = H_V(m) - H_V(m \mid D_{t_1}) \approx 0.626; \\
\mathrm{VIG}(m, D_{t_2}; q = 1) & = H_V(m) - H_V(m \mid D_{t_2}) \approx 0.575,
\end{align}
which matches our intuition that $D_{t_1}$ offers a higher IG than $D_{t_2}$.
We also see that both VIG values are less than MI, indicating that VIG offers a more conservative quantification of IG due to the similarity between outcomes $m = A$ and $m = B$.

\begin{figure}[t]
\centering
\includegraphics[width=\linewidth]{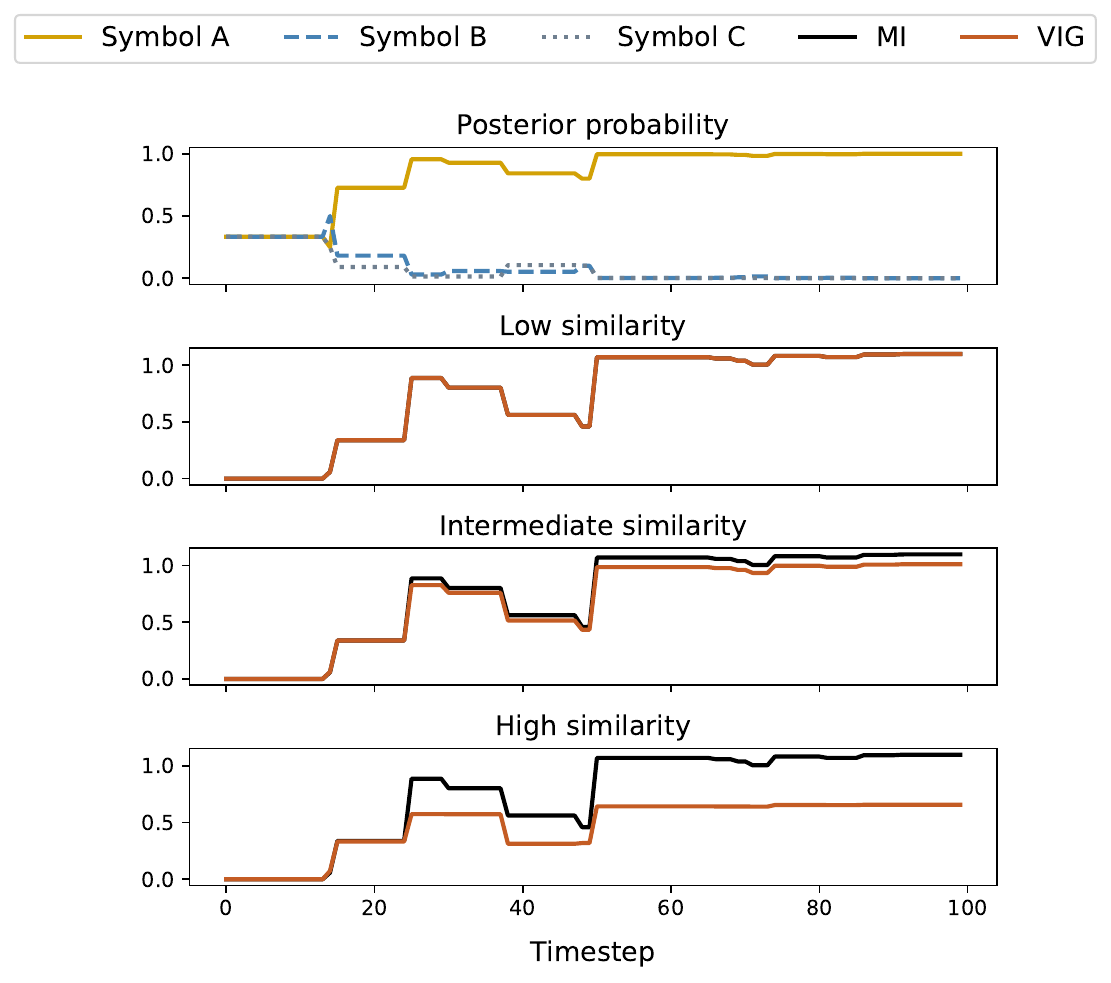}
\caption{
Progression of posterior probabilities and information gain, as measured by MI and VIG, under varying degrees of inter-message similarity as a function of time.
MI plots the same trend across different conditions, while VIG becomes lower under more similar messages, appropriately quantifying increasing difficulty in distinguishing similar messages.
}
\label{fig:neuro_progress}
\end{figure}

In \cref{fig:neuro_progress}, we perform the simulation with the model by \citet{christie2023information} by transmitting three symbols $A$, $B$, and $C$ with varying degrees of similarity.
In particular, we construct the kernel matrix:
\begin{equation}
K = \begin{bmatrix}
1 & s & 0 \\
s & 1 & 0 \\
0 & 0 & 1
\end{bmatrix},
\end{equation}
where $C$ is modeled as being distinct from $A$ and $B$, while the similarity between $A$ and $B$ varies such that
\begin{itemize}
\item the covariance $s = 0$ in the low-similarity setting,
\item the covariance $s = 0.5$ in the intermediate-similarity setting, and
\item the covariance $s = 0.9$ in the high-similarity setting.
\end{itemize}

Symbol $A$ is set to be the intended stimulus, and the top panel of \cref{fig:neuro_progress} shows the posterior probability that each symbol is the intended stimulus $p(m \mid D_t)$ as the function of time $t$.
Here, we start out with a flat prior where each symbol has a $1 / 3$ probability of being the intended stimulus, and as we observe more data as time goes on, $p(m = A \mid D_t)$ increases, indicating that $m = A$ is the true message.

The remaining panels of \cref{fig:neuro_progress} show the progression of IG as measured by MI and VIG in each of the three aforementioned settings.
MI treats individual stimuli as completely distinct from one another and plots the same trend across the settings, failing to distinguish the different conditions.
Meanwhile, VIG appropriately quantifies increasing difficulty in distinguishing messages as they become more similar, as demonstrated by lower IG.
(In the extreme case of low similarity where there is no inter-stimulus similarity, VIG reduces to MI.)

\subsection{The active data acquisition framework}

In traditional active data acquisition, we choose the data point that possesses the largest MI with $\theta$:
\begin{equation}
\label{eq:ig_bed}
\begin{split}
x_* = & \argmax_x \Big( H(\theta) - H(\theta \mid x) \Big) \\
= & \argmax_x \Big( H(\theta) - \mathbb{E}_{p(y \mid x, \theta)} [H(\theta \mid x, y)] \Big),
\end{split}
\end{equation}
where we have made the dependence on the observed data implicit.

We use the following the criterion for VIG:
\begin{equation}
\label{eq:vig_bed}
x_* = \argmax_x \Big( H_V(D; q) - \mathbb{E}_{p(y \mid x, \theta)} [H_V(D_y \mid x, y; q)] \Big),
\end{equation}
where, similarly to \cref{eq:vig}, $D$ and $D_y$ denote the set of prior and posterior samples of $\theta$ (conditioned on each realization of $y$ in the latter case).
As $H_V(D)$ is the Vendi entropy of the prior samples of $\theta$ given the current data and therefore constant across different designs $x$, this VIG criterion becomes:
\begin{equation}
\label{eq:vig_bed_alt}
x_* = \argmin_x \mathbb{E}_{p(y \mid x, \theta)} [H_V(D_y \mid x, y; q)].
\end{equation}
We now describe the tasks shown in \cref{fig:bed_results} in more details.
\paragraph{Step functions.}
Consider the task of learning a step function $f \colon [0, 1] \mapsto \{ 0, 1 \}$ parameterized by an unknown threshold $\theta \in [0, 1]$ such that $y = f(x) = \mathbb{I} [x \geq \theta]$,
where $\mathbb{I}[\cdot]$ is the indicator function. 
The prior $p(\theta)$ is constructed to be a mixture of multiple Beta distributions with randomized parameters.
The resulting multimodality in $p(\theta)$ induces a more complicated problem than under $U[0, 1]$, which can be solved optimally with binary search.
\paragraph{Death process.}
Modeling commonly faced problems in epidemiology, the observation model by \citet{cook2008optimal} assumes that healthy individuals become infected at unknown rate $\theta$, and we want to choose observation times $x$ at which to observe the number of infected individuals $y$.
Our goal is to gather data to best infer the unknown infection rate $\theta$ between $0$ and $10$.
\paragraph{Location sensing.}
\citet{foster2021deep} proposed a problem inspired by the acoustic energy attenuation model of \citet{sheng2004maximum} where we want to identify the locations of two hidden energy sources, each emitting a signal whose intensity attenuates according to the inverse-square law.
At each location within the search space we choose to inspect, we can measure the total intensity equal to the sum of the intensity levels of the two signals.

At each iteration of the loop, we measure the absolute error between the current \emph{maximum a posteriori} (MAP) estimate of $\theta$ and its true value for the first two tasks.
In location sensing, we compute the sum of pairwise Euclidean distances between the MAP estimate and the two unknown locations, selecting the pairing that yields the lower error,.
We set the experimentation budget $B = 5$ for the step function learning task, and $B = 20$ in the other two.
Each run is repeated 10 times with a new value of $\theta$ sampled at random, and we include random sampling across the search space as a search baseline.

We now describe the LSE model.
Assume we are given a discrete pool of points $\mathcal{X}~=~\{ x_i \}_{i = 1}^N$ that make up the search space.
The behavior of these data points is governed by a latent function $f$, whose value $y_i = f(x_i)$ at an input $x_i \in \mathcal{X}$ of our choosing may be obtained; we call $y_i$ the label of the data point $x_i$.
We are interested in whether each label exceeds a certain threshold $h$, concatenated into a binary vector $\boldsymbol{\theta}$:
\begin{equation}
\label{eq:theta_lse}
\boldsymbol{\theta} =
\begin{bmatrix}
\mathbb{I} [y_1 > h] \\
\mathbb{I} [y_2 > h] \\
\vdots \\
\mathbb{I} [y_N > h]
\end{bmatrix},
\end{equation}
where $\mathbb{I}[\cdot]$ is the indicator function.

To perform LSE, we first build a predictive model for the function $f$ trained on $\mathcal{D}$, the points whose labels have been observed so far throughout the iterative procedure.
A flexible choice for this model is a Gaussian process (GP), which induces a Gaussian belief about the label $y$ of any given point $x$:
\begin{equation}
\label{sec:normal}
p(y \mid x) = \mathcal{N} \big( \mu(x), K(x, x) \big),
\end{equation}
where $\mu$ and $K$ are the mean function and kernel of the GP, respectively; we refer to \citet{rasmussen2006gaussian} for more details on GPs.

To design a method for identifying informative data, the traditional approach is to minimize the entropy of $\boldsymbol{\theta}$, that is, find the data point $x$ whose label $y$ has the highest mutual information with $\boldsymbol{\theta}$ on average:
\begin{equation}
\begin{split}
x_* & = \argmin_{x \in \mathcal{X}} ~ \mathbb{E}_y \! \Big[ H \! \big( \boldsymbol{\theta} \mid \mathcal{D} \cup \{ x, y \} \big) \Big] \\
& = \argmax_{x \in \mathcal{X}} ~ H \! \big( \boldsymbol{\theta} \mid \mathcal{D} \big) - \mathbb{E}_y \! \Big[ H \! \big( \boldsymbol{\theta} \mid \mathcal{D} \cup \{ x, y \} \big) \Big] \\
& = \argmax_{x \in \mathcal{X}} ~ I(\boldsymbol{\theta}; y \mid x, \mathcal{D}).
\end{split}
\end{equation}
The widely used STRADDLE policy by \citet{bryan2005active} targets the location $x$ we are the most uncertain about whether the corresponding label $y$ exceeds the threshold $h$, minimizing $| \mu(x) - h |$.
To avoid repeatedly querying a boundary region and consequently getting stuck, \citet{bryan2005active} added an exploration term in the form of the predictive standard deviation $\sigma(x)$ to encourage inspecting regions far away from observed data, and proposed the criterion:
\begin{equation}
\label{eq:straddle}
x_* = \argmax_{x \in \mathcal{X}} 1.96 \, \sigma(x) - | \mu(x) - h |.
\end{equation}

To realize VIG for LSE, we require a kernel that compares two given binary vectors $\boldsymbol{\theta}_1$ and $\boldsymbol{\theta}_2$; we choose to normalize and invert the Hamming distance such that:
\begin{equation}
\label{eq:lse_kernel}
k(\boldsymbol{\theta}_1, \boldsymbol{\theta}_2) = 1 - \frac{d_H(\boldsymbol{\theta}_1, \boldsymbol{\theta}_2)}{N},
\end{equation}
where $d_H$ denotes the Hamming distance between the two input vectors, which outputs the number of mismatches between them, and $N$ is the length of each vector.
In other words, this kernel computes the level of agreement between two samples proportional to their size.

In our experiments, each run starts out with one observed point selected uniformly at random from the pool $\mathcal{X}$, which then repeats ten times.

\section{Additional results}
\label{sec:add_results}

We now present results from ablation studies to investigate the impacts of the order $q$ in \cref{eq:vs_q} of the VIG policy, which was set at $q = 1$ in the main paper.
As discussed in \citet{pasarkar2024cousins}, the hyperparameter $q$ affects the sensitivity of the metric to rarity: low values of $q$ leads to more sensitivity to rare features of the items, while high values of $q$ prioritize common features of the items.
By setting this hyperparameter, we can induce a family of VIG criteria with different levels of sensitivity to rare samples.

\begin{figure}[t]
\includegraphics[width=0.94\linewidth]{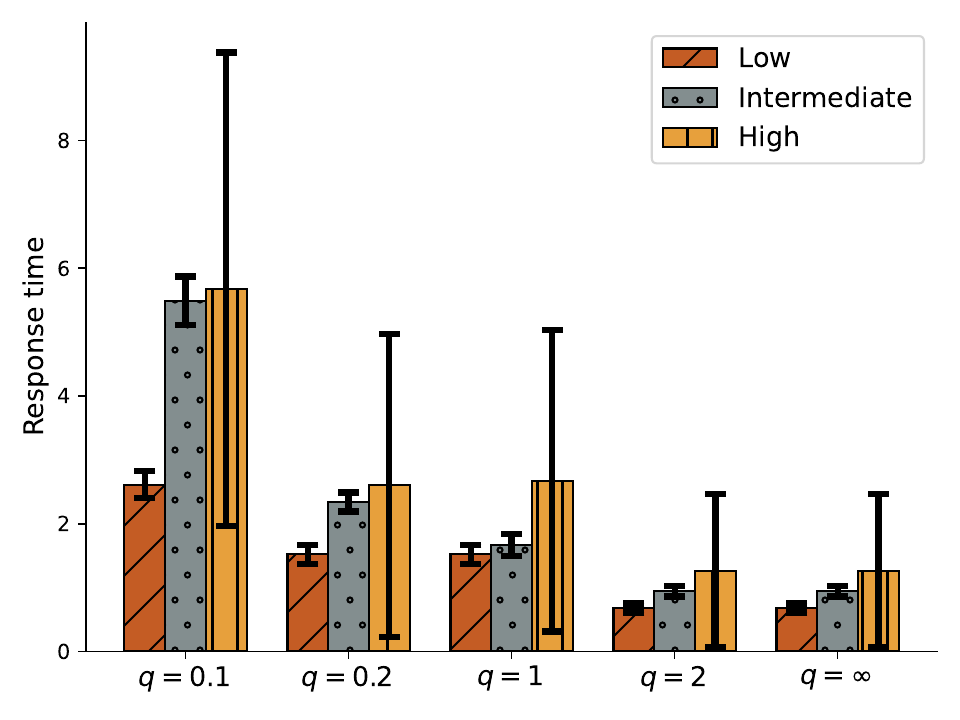}
\caption{
Simulated response times as a function of similarity between possible messages by VIG of different orders $q$.
}
\label{fig:neuro_compare_q}
\end{figure}

\begin{figure*}[t]
\centering
\includegraphics[width=\linewidth]{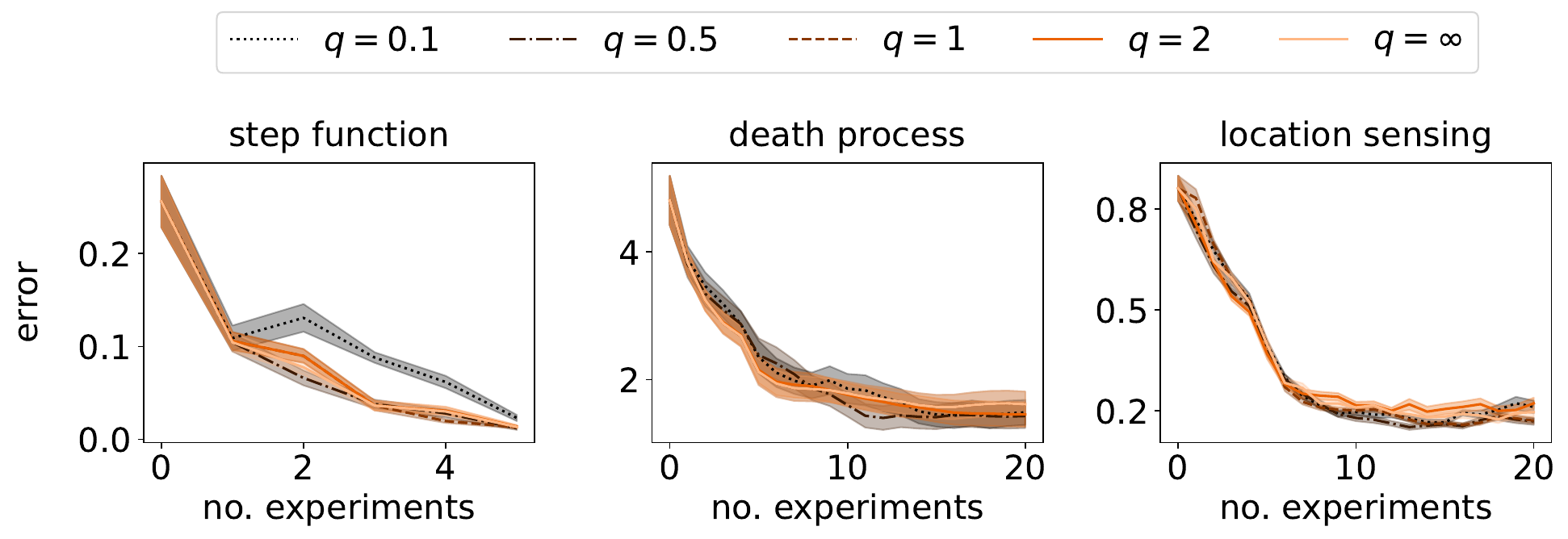}
\caption{
Estimation errors achieved by VIG of different orders $q$ as a function of the number of queries.
VIG's performance is mostly robust against the value of $q$.
}
\label{fig:bed_q}
\end{figure*}

\begin{figure*}[t]
\centering
\includegraphics[width=\linewidth]{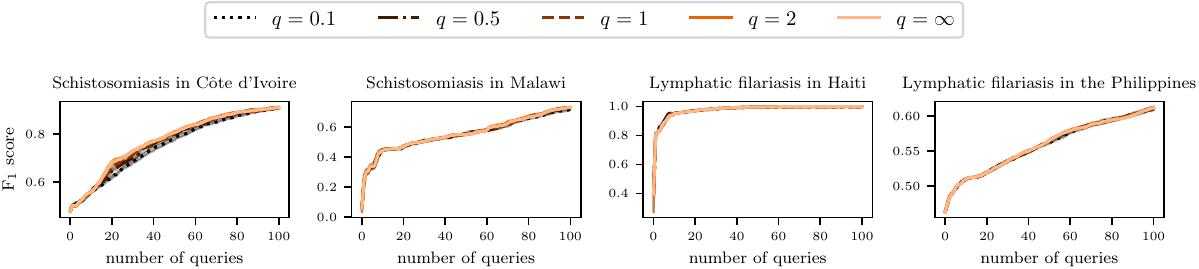}
\caption{
F$_1$ scores achieved by VIG of different orders $q$ as a function of the number of queries.
VIG's performance is mostly robust against the value of $q$.
}
\label{fig:lse_q}
\end{figure*}

\begin{table}[t]
\centering
\caption{
Quantification of information gain about a channel's input upon observing its output.
\textbf{First column}: VIG reduces to MI under transmitted symbols that are dissimilar from one another.
\textbf{Second and third column}: MI considers the middle and last channels to have equal capacity, as each output corresponds to the same number of input values; VIG on the other hand prioritizes the last channel, agreeing with the mean absolute error when predicting the transmitted input.
}
\label{tab:channel_v2}
\begin{tabular}{cccc}
\toprule
& \begin{tikzpicture}
    
    
    

    \node (x) at (0,2) {$x$};
    \node (y) at (0,0) {$y$};
    \node (z) at (0,-2) {$z$};
    
    \node (a) at (3,1) {$a$};
    \node (b) at (3,-1) {$b$};
    
    \foreach \a in {a, b} {
        \foreach \x in {x, y, z} {
            \draw[->] (\x) -- (\a);
        }
    }
    

\end{tikzpicture} & \begin{tikzpicture}
    \node (a) at (3,1.5) {$a$};
    \node (b) at (3,0.5) {$b$};
    \node (c) at (3,-0.5) {$c$};
    \node (d) at (3,-1.5) {$d$};
    
    \node (x1) at (0,2) {$0$};
    \node (x2) at (0,1.2) {$0.1$};
    \node (x3) at (0,0.4) {$0.2$};
    \node (x4) at (0,-0.4) {$0.3$};
    \node (x5) at (0,-1.2) {$0.4$};
    \node (x6) at (0,-2) {$0.5$};
    
    \draw[->] (x1) -- (a);
    \draw[->] (x1) -- (c);
    \draw[->] (x2) -- (b);
    \draw[->] (x2) -- (d);
    \draw[->] (x3) -- (a);
    \draw[->] (x3) -- (c);
    \draw[->] (x4) -- (b);
    \draw[->] (x4) -- (d);
    \draw[->] (x5) -- (a);
    \draw[->] (x5) -- (c);
    \draw[->] (x6) -- (b);
    \draw[->] (x6) -- (d);
\end{tikzpicture} & \begin{tikzpicture}
    \node (a) at (3,1.5) {$a$};
    \node (b) at (3,0.5) {$b$};
    \node (c) at (3,-0.5) {$c$};
    \node (d) at (3,-1.5) {$d$};
    
    \node (x1) at (0,2) {$0$};
    \node (x2) at (0,1.2) {$0.1$};
    \node (x3) at (0,0.4) {$0.2$};
    \node (x4) at (0,-0.4) {$0.3$};
    \node (x5) at (0,-1.2) {$0.4$};
    \node (x6) at (0,-2) {$0.5$};
    
    \draw[->] (x1) -- (a);
    \draw[->] (x1) -- (b);
    \draw[->] (x2) -- (a);
    \draw[->] (x2) -- (b);
    \draw[->] (x3) -- (a);
    \draw[->] (x3) -- (b);
    \draw[->] (x4) -- (c);
    \draw[->] (x4) -- (d);
    \draw[->] (x5) -- (c);
    \draw[->] (x5) -- (d);
    \draw[->] (x6) -- (c);
    \draw[->] (x6) -- (d);
\end{tikzpicture} \\
\midrule
$\mathrm{MI}$ & $0.176$ & $0.693$ & $0.693$ \\
$\mathrm{VIG}$ & $0.176$ & $0.014$ & $0.146$ \\
\midrule
MAE & -- & 0.178 & 0.089 \\
\bottomrule
\end{tabular}
\end{table}

First, \cref{fig:neuro_compare_q} shows the distribution of response times by VIG under different orders $q$, where larger values of $q$ lead to shorter response times.
This trend is explained by the effects of $q$ mentioned above: as small orders are more sensitive to rarity, more data, and thus more time, is required for IG about the most common class to exceed the threshold.
\cref{fig:bed_q,fig:lse_q} show the results of VIG across a wide range of values for $q$ under active data acquisition tasks.
We observe the reassuring trend that different versions of VIG behave similarly, demonstrating the VIG's robustness against this hyperparameter.

\end{document}